\documentclass{aa}
\usepackage{natbib}
\newcount\longrefs
\def\adaa{\ifnum\longrefs=1 {Advances in Astronomy and Astrophysics}\else 
                           {Adv. Astron.\ Astrophys.}\fi}
\def\aap{\ifnum\longrefs=1 {Astron.\ Astrophys.}\else 
                           {A\hbox{\rm \&}A}\fi}
\def\aapr{\ifnum\longrefs=1 {Astron.\ Astrophys.\ Rev.}\else 
                            {A\hbox{\rm \&}AR}\fi}
\def\aaps{\ifnum\longrefs=1 {Astron.\ Astrophys.\ Suppl.}\else 
                            {A\hbox{\rm \&}AS}\fi}
\def\aj{\ifnum\longrefs=1 {Astron.\ J.}\else 
                          {AJ}\fi} 
\def\ao{\ifnum\longrefs=1 {Applied Optics}\else 
                           {Appl.\ Opt.}\fi} 
\def\aspcs{\ifnum\longrefs=1 {Astron.\ Soc.\ Pacific Conf. Series}\else 
                           {ASP Conf.\ Ser.}\fi} 
\def\apj{\ifnum\longrefs=1 {Astrophys.\ J.}\else 
                           {ApJ}\fi} 
\def\apjl{\ifnum\longrefs=1 {Astrophys.\ J. Lett.}\else 
                            {ApJ}\fi} 
\def\aplett{\ifnum\longrefs=1 {Astrophys.\ J. Lett.}\else 
                            {ApJ}\fi} 
\def\apjs{\ifnum\longrefs=1 {Astrophys.\ J. Suppl.}\else 
                            {ApJS}\fi}
\def\apss{\ifnum\longrefs=1 {Astrophys.\ and Space Science}\else 
                            {Ap\&SS}\fi}
\def\araa{\ifnum\longrefs=1 {Ann.\ Rev.\ Astron.\ Astrophys.}\else 
                            {ARA\hbox{\rm \&}A}\fi}
\def\azh{\ifnum\longrefs=1 {Astronomicheskii Zhurnal}\else 
                            {Astron.\ Zhur.}\fi}
\def\baas{\ifnum\longrefs=1 {Bull.\ Am.\ Astron.\ Soc.}\else 
                            {BAAS}\fi}
\def\bain{\ifnum\longrefs=1 {Bull.\ Astronom.\ Institutes Netherlands}\else
                            {Bull.\ Astr.\ Inst.\ Neth.}\fi}
\def\gca{\ifnum\longrefs=1 {Geochim.\ Cosmochim.\ Acta}\else 
                           {Geochim.\ Cosmochim.\ Acta}\fi}
\def\geo{\ifnum\longrefs=1 {Geophysical Journal}\else 
                           {Geophys.\ J.}\fi}
\def\grl{\ifnum\longrefs=1 {Geophys.\ Res.\ Lett.}\else 
                           {Geoph.\ Res.\ Lett.}\fi}
\def\iaucirc{\ifnum\longrefs=1 {IAU Circulars}\else 
                          {IAU Circ.}\fi}
\def\ip{\ifnum\longrefs=1 {in press}\else 
                          {in press}\fi}
\def\jgr{\ifnum\longrefs=1 {J.\ Geophys.\ Res.}\else 
                           {J.\ Geophys.\ Res.}\fi}  
\def\jrasc{\ifnum\longrefs=1 {J.\ Royal Astron.\ Soc.\ Canada}\else 
                           {JRAS Can.}\fi}  
\def\mnras{\ifnum\longrefs=1 {Mon.\ Not.\ Roy.\ Astron.\ Soc.}\else 
                             {MNRAS}\fi} 
\def\nat{\ifnum\longrefs=1 {Nature}\else 
                           {Nat}\fi}
\def\pasj{\ifnum\longrefs=1 {Pub.\ Astron.\ Soc.\ Japan}\else 
                            {PASJ}\fi} 
\def\pasp{\ifnum\longrefs=1 {Pub.\ Astron.\ Soc.\ Pacific}\else 
                            {PASP}\fi} 
\def\physscr{\ifnum\longrefs=1 {Physica Scripta}\else 
                            {Phys.\ Scrip.}\fi} 
\def\planss{\ifnum\longrefs=1 {Planetary \& Space Science}\else 
                            {Plan. \& Space Sci.}\fi} 
\def\procspie{\ifnum\longrefs=1 {Proc.\ SPIE}\else 
                            {Proc.\ SPIE}\fi} 
\def\qjras{\ifnum\longrefs=1 {Quarterly J.\ Royal Astron.\ Soc.}\else 
                            {QJRAS}\fi} 
\def\sa{\ifnum\longrefs=1 {Soviet Astron..}\else 
                               {Sov.\ Astron.}\fi}
\def\skytel{\ifnum\longrefs=1 {Sky \& Telescope}\else 
                            {Sky \& Tel.}\fi} 
\def\solphys{\ifnum\longrefs=1 {Solar Phys.}\else 
                               {Solar Phys.}\fi}
\def\ssr{\ifnum\longrefs=1 {Space Science Rev.}\else 
                               {Space\ Sci.\ Rev.}\fi}

\def\bibfiles{/home/bwillems/latex/bibtex/bibliofile}   

\def\aareferences{\longrefs=0  \bibliographystyle{/home/bwillems/latex/bibtex/aabib}
             \bibliography{/home/bwillems/latex/bibtex/aajour,\bibfiles}}

\def\dutch{\def\refname{Referenties}\def\abstractname{Samenvatting}%
  \def\bibname{Bibliografie}\def\chaptername{Hoofdstuk}%
  \def\appendixname{Bijlage}\def\contentsname{Inhoudsopgave}%
  \def\listfigurename{Lijst van figuren}\def\listtablename{Lijst van tabellen}%
  \def\indexname{Index}\def\figurename{Figuur}\def\tablename{Tabel}%
  \def\partname{Deel}\def\enclname{Bijlage(n)}\def\ccname{Ter attentie van}%
  \def\headtoname{Aan}\def\headpagename{Pagina}%
  \def\today{\number\day\space\ifcase\month\or januari\or februari\or maart\or%
     april\or mei\or juni\or juli\or augustus\or september\or oktober\or%
     november\or december\fi \space\number\year}%
  \typeout{
              >>>>> use hlatex209 for Dutch hyphenation <<<<< 
         }}
\newcounter{onefig} \newcounter{fignumber}
\newcount\nocaptions \newcount\nofigures \newcount\figwidth
\newcount\viewgraphs
  \def\paper{}  \def\figlabel{} 
\long\def\nextfig#1{\setcounter{figure}{\value{fignumber}}
  \addtocounter{fignumber}{1}
  \ifnum \viewgraphs=1 \newpage \pagestyle{empty} \fi 
  \ifnum\value{onefig}=0 #1 \fi                 
  \ifnum\value{onefig}=\value{fignumber} #1 \fi}
\def\figwidths#1#2{\ifnum \nocaptions=1 #2mm \else #1mm \fi}  
\def\paper#1{}  
\long\def\plotfig#1#2{\ifnum \nofigures=1 \else #2 \fi}
\long\def\captiontext#1{\ifnum \nofigures=1 \raggedright \fi 
   \ifnum \nocaptions=1 \paper
     \ifnum \viewgraphs=0 
       \newline  \mbox{}\hrulefill\mbox{} \newline 
       \newline label:~\{\figlabel\} 
     \fi 
     \else \ifnum \nofigures=0 \fi 
   #1 \fi}

\newcount\panelwidth \newcount\panelheight 
\newcount\bxmin \newcount\bymin \newcount\bxmax \newcount\bymax
\newcount\tbxmin \newcount\tbymin
\newcount\tpanelwidth \newcount\tpanelheight \newcount\tpdif
\def\panelsize #1,#2;{\panelwidth=#1 \panelheight=#2}  
\def\setbb #1,#2;#3,#4;#5,#6;{
  \tbxmin=#1 \tbymin=#2    
  \bxmin=#3 \bymin=#4      
  \bxmax=#5 \bymax=#6}     
\def\barepanel #1{%
  \ifnum\panelheight=0 
    \tpdif=\bymax \advance\tpdif by -\bymin
    \multiply \tpdif by \panelwidth
    \tpanelheight=\tpdif
    \tpdif=\bxmax \advance\tpdif by -\bxmin
    \divide \tpanelheight by \tpdif
  \else \tpanelheight=\panelheight \fi
  \epsfig{file=#1,%
     bbllx=\bxmin bp,bblly=\bymin bp,bburx=\bxmax bp,bbury=\bymax bp,clip=,%
     width=\panelwidth mm,height=\tpanelheight mm}}
\def\labelypanel #1{
  \ifnum\panelheight=0 
    \tpdif=\bymax \advance\tpdif by -\bymin
    \multiply \tpdif by \panelwidth
    \tpanelheight=\tpdif
    \tpdif=\bxmax \advance\tpdif by -\bxmin
    \divide \tpanelheight by \tpdif
  \else \tpanelheight=\panelheight \fi
  \tpdif=\bxmax \advance\tpdif by -\tbxmin
  \tpanelwidth=\panelwidth \multiply \tpanelwidth by \tpdif
  \tpdif=\bxmax \advance\tpdif by -\bxmin
  \divide \tpanelwidth by \tpdif
  \epsfig{file=#1,%
    bbllx=\tbxmin bp,bblly=\bymin bp,bburx=\bxmax bp,bbury=\bymax bp,%
    clip=,width=\tpanelwidth mm,height=\tpanelheight mm}}
\def\labelxpanel #1{%
  \ifnum\panelheight=0 
    \tpdif=\bymax \advance\tpdif by -\bymin
    \multiply \tpdif by \panelwidth
    \tpanelheight=\tpdif
    \tpdif=\bxmax \advance\tpdif by -\bxmin
    \divide \tpanelheight by \tpdif
  \else \tpanelheight=\panelheight \fi
  \tpdif=\bymax \advance\tpdif by -\tbymin
  \multiply \tpanelheight by \tpdif
  \tpdif=\bymax \advance\tpdif by -\bymin
  \divide \tpanelheight by \tpdif
  \epsfig{file=#1,%
    bbllx=\bxmin bp,bblly=\tbymin bp,bburx=\bxmax bp,bbury=\bymax bp,%
    clip=,width=\panelwidth mm,height=\tpanelheight mm}}
\def\labelxypanel #1{%
  \ifnum\panelheight=0 
    \tpdif=\bymax \advance\tpdif by -\bymin
    \multiply \tpdif by \panelwidth
    \tpanelheight=\tpdif
    \tpdif=\bxmax \advance\tpdif by -\bxmin
    \divide \tpanelheight by \tpdif
  \else \tpanelheight=\panelheight \fi
  \tpdif=\bxmax \advance\tpdif by -\tbxmin
  \tpanelwidth=\panelwidth \multiply \tpanelwidth by \tpdif
  \tpdif=\bxmax \advance\tpdif by -\bxmin
  \divide \tpanelwidth by \tpdif 
  \tpdif=\bymax \advance\tpdif by -\tbymin 
  \multiply \tpanelheight by \tpdif
  \tpdif=\bymax \advance\tpdif by -\bymin
  \divide \tpanelheight by \tpdif
  \epsfig{file=#1,%
    bbllx=\tbxmin bp,bblly=\tbymin bp,bburx=\bxmax bp,bbury=\bymax bp,%
    clip=,width=\tpanelwidth mm,height=\tpanelheight mm}}

\def\CC{\par \vspace*{-2ex} \footnotesize \baselineskip=8pt \begin{verbatim}}

\long\def\startignore #1\stopignore{}   

\def\setlistparams{         
  \topsep=0.7ex                 
  \itemsep=0.7ex                
  \leftmargini=3ex}             
\setlistparams                  

\newcounter{alistindex}       

\newcounter{romenumnr}

\newlength{\minipagewidth}

\newsavebox{\boxcontent}
\newcommand{\ovalhead}[1]{
  \unitlength=1cm
  \sbox{\boxcontent}{\mbox{~~{#1}~~}}
  \begin{center}
    \ifdim\wd\boxcontent>6ex 
    \ifdim\wd\boxcontent<8cm 
    \begin{picture}(8,3) \thicklines     
      \put(4.0,0.8){\oval(8,1.6)} 
      \put(0.0,0.7){\parbox{8cm}{
         \begin{center} \usebox{\boxcontent} \end{center}}}
    \end{picture}
    \else \ifdim\wd\boxcontent<12cm 
    \begin{picture}(12,3) \thicklines     
        \put(6.0,0.8){\oval(12,1.6)} 
        \put(0.0,0.7){\parbox{12cm}{
           \begin{center} \usebox{\boxcontent} \end{center}}}
    \end{picture}
    \else
    \begin{picture}(16,3) \thicklines     
        \put(8.0,0.8){\oval(16,1.6)} 
        \put(0.0,0.7){\parbox{16cm}{
           \begin{center} \usebox{\boxcontent} \end{center}}}
    \end{picture}
    \fi \fi \fi
  \end{center}} 

\newcounter{headnr}            
\newcounter{subheadnr}[headnr]
\newcounter{subsubheadnr}[subheadnr]
\def\head #1\par{
  \stepcounter{headnr}                          
  \vspace{2ex} \noindent                        
  {\bf \theheadnr~~~~#1}\\[1ex] \noindent}      
\def\subhead #1\par{  
  \stepcounter{subheadnr}
  \vspace{1.3ex} \noindent
  {\bf \theheadnr.\arabic{subheadnr}~~~#1}\\[0.3ex] \noindent}
\def\subsubhead #1\par{
  \stepcounter{subsubheadnr}
  \vspace{1.0ex} \noindent
  {\bf \theheadnr.\arabic{subheadnr}.\arabic{subsubheadnr}~~~#1}\\ \noindent}

\font\dropfont= cmr12 scaled \magstep5
\def\dropcap#1#2{{\noindent
    \setbox0\hbox{\dropfont #1}\setbox1\hbox{#2}\setbox2\hbox{(}%
    \count0=\ht0\advance\count0 by\dp0\count1\baselineskip
    \advance\count0 by-\ht1\advance\count0by\ht2
    \dimen1=.5ex\advance\count0by\dimen1\divide\count0 by\count1
    \advance\count0 by1\dimen0\wd0
    \advance\dimen0 by.25em\dimen1=\ht0\advance\dimen1 by-\ht1
    \global\hangindent\dimen0\global\hangafter-\count0
    \hskip-\dimen0\setbox0\hbox to\dimen0{\raise-\dimen1\box0\hss}%
    \dp0=0in\ht0=0in\box0}#2}

\def\level #1 #2#3#4{$#1 \: ^{#2} \mbox{#3} ^{#4}$}   

\def\mathstacksym#1#2#3#4#5{\def#1{\mathrel{\hbox to 0pt{\lower 
    #5\hbox{#3}\hss} \raise #4\hbox{#2}}}}

\mathstacksym\lta{$<$}{$\sim$}{1.5pt}{3.5pt} 
\mathstacksym\gta{$>$}{$\sim$}{1.5pt}{3.5pt} 
\mathstacksym\lrarrow{$\leftarrow$}{$\rightarrow$}{2pt}{1pt} 
\mathstacksym\lessgreat{$>$}{$<$}{3pt}{3pt} 

\usepackage{graphicx}
\bibpunct{(}{)}{;}{a}{}{,}

\begin{document} 
\title{Tidally induced radial-velocity variations in close binaries} 
\author{B.\ Willems\inst{1} \and C.\ Aerts\inst{2}} 
\institute{Department of Physics and Astronomy, The Open University,
Walton Hall, Milton Keynes, MK7 6AA, UK \\ 
\email{B.Willems@open.ac.uk}
\and 
Instituut voor Sterrenkunde, Katholieke Universiteit Leuven, 
Celestijnenlaan 200\,B, B-3001 Leuven, Belgium \\
\email{Conny.Aerts@ster.kuleuven.ac.be}} 
\date{Received date; accepted date} 
 
\abstract{A theoretical framework for the determination of tidally
  induced radial-velocity variations in a component of a close binary
  is presented. Both the free and the forced oscillations of the
  component are treated as linear, isentropic perturbations of a
  spherically symmetric star. Resonances between dynamic
  tides and free oscillation modes are taken into account by means of
  the formalism developed by \citet{SWV1998}. The amplitude of the
  tidally induced radial-velocity variations seen by the observer
  depends on the orbital eccentricity and on the orbital inclination.  
  The amplitude increases with increasing orbital eccentricity and is
  most sensitive to the value of the orbital inclination when
  $20^\circ \la i \la 70^\circ$. In the case of a $5\,M_\odot$ ZAMS
  star with a $1.4\,M_\odot$ compact companion, it is shown that
  resonant dynamic tides can lead to radial-velocity variations with
  amplitudes large enough to be detected in observations. The shape of
  the tidally induced radial-velocity curves varies from very irregular for
  orbital periods away from any resonances with free oscillation modes
  to sinusoidal for orbital periods close to a resonance with 
  a free oscillation mode. Our investigation is concluded with an
  application to the slowly pulsating B star HD\,177863 showing the
  possibility of resonant excitation of a high-order second-degree
  $g^+$-mode in this  star.  
}

\maketitle

\section{Introduction}

In close binary systems of stars, each component is subject to the
time-dependent tidal force exerted by its companion. A commonly used
approach for the study of tidal effects in close binaries is based on
the expansion of the tide-generating potential in Fourier series in
terms of multiples of the companion's mean motion. Through these
expansions, the tidal action of the companion induces an infinite
number of partial dynamic tides in the star, each with its own forcing
angular frequency.

In binaries with shorter orbital periods, the forcing angular
frequencies may be close to the eigenfrequencies of the free
oscillation modes $g^+$ of the component stars \citep{Cow1941}. These
proximities lead to resonances which enhance the tidal motions of
the mass elements and can have significant consequences for the
observational properties of the binary. 

\citet{Zahn1970} studied resonances of dynamic tides with
low-frequency $g^+$-modes in non-rotating stars consisting of a
convective core and a radiative envelope. For his purpose, he used an
asymptotic representation for the eigenfunctions of the $g^+$-modes
which he established neglecting the perturbation of the gravitational
field due to the star's tidal distortion. Zahn's treatment was later
generalised by \citet{Roc1987} to include the effects of the Coriolis
force in slowly rotating binary components.

The luminosity variations associated with tidally excited oscillation 
modes in close binaries were investigated by \citet{Kum1995}. The
authors proposed a new method for the determination of the orbital
inclination based on fitting the {\it shape} of theoretically derived light
curves to the shape of observationally determined light curves. They
applied their 
results to various polytropic stellar models
and to the binary pulsars PSR\,J0045-7319 and PSR\,B1259-63.

More recently, \citet{SWV1998} (hereafter referred to as Paper~I)
derived semi-analytical solutions for the components of a resonant
dynamic tide by means of a two-time variable expansion procedure 
in which both the free and the forced oscillations of the star were
treated as linear, isentropic oscillations of a spherically symmetric
star. 
The authors concluded that, at the lowest-order of approximation, the
oscillation mode involved in the resonance is excited with the forcing
frequency of the resonant dynamic tide.  Their investigation
was subsequently extended by \citet{W2001} to take into account the
effects of radiative damping in the nonadiabatic surface layers of a
star. 

The duration of a resonance is determined by the combined effect of
stellar and orbital evolution. \citet{Wit1999b,Wit2001} have shown
that when both effects are taken into account, a dynamic tide can
easily become locked in a resonance for a prolonged period of
time. Such a long-term resonance can 
have a significant influence on the secular evolution of the orbital
elements and increases the probability of detecting the tidally
induced oscillations in the star.

From an observational point of view, firm evidence of tidally excited
oscillations is still scarce, possibly due to the absence of any
systematic observational study on the nature of the pulsations
discovered in close binaries. Studies of individual stars are reported
in the literature, e.g.\ $\sigma\,$Sco \citep{Fitch1967}, 14\,Aur\,A
\citep{Fitch1979}, Spica \citep{Smith1985a,Smith1985b}, 16\,Lac
\citep{Chap1995}, V\,539\,Arae \citep{Claus1996}, among others, but
the resonant nature of the modes was hardly ever proven.  An
introduction to a systematic observational study in the case of
early-type binaries is given by \citet{Aerts1998}. Their initiative
resulted from a more general long-term systematic study of
line-profile variations in $\beta\,$Cep stars and in slowly pulsating
B stars, which led to the assessment that many of these types of
pulsators belong to a close binary. Further evidence for the presence
of B-type $g$-mode pulsators in close binaries was found by
\citet{DeCat2000}. An additional systematic observational project in
this respect is the SEarch for FOrced Nonradial Oscillations (SEFONO)
by \citet{Harm1997}.

In this investigation, our aim is to determine the radial-velocity
variations associated with resonantly excited oscillation modes in
binaries with short orbital periods. To this end, we use
semi-analytical solutions for the components of a resonant dynamic
tide derived by \citet{SWV1998}. We conclude our investigation by
applying our results to the slowly pulsating B star HD\,177863.

The plan of the paper is as follows. In Sects.~2 and~3, we present the
basic assumptions adopted in our investigation and we decompose the
tide-generating potential in terms of spherical harmonics and in
Fourier series in terms of the companion's mean motion. In Sects.~4
and~5, we derive an expression for the radial-velocity variations
associated with a resonant dynamic tide. 
The contributions of non-resonant dynamic tides to the tidally induced
radial-velocity variations are determined in Sect.~6.  In Sect.~7, we
determine the total variation of a star's radial
velocity due to the various resonant and non-resonant dynamic tides
and we illustrate the role of the orbital eccentricity and the orbital
inclination. In Sect.~8, we apply our results to a binary consisting
of a $5\,M_\odot$ zero-age main sequence star and a $1.4\,M_\odot$
compact companion. In Sect.~9, the possibility of a
resonantly excited oscillation mode in the slowly pulsating B
star HD\,177863 is investigated. The final section is devoted to
concluding remarks.

\section{Basic assumptions}

\label{basic}

Consider a close binary system of stars that are orbiting around each
other in an unvarying Keplerian orbit with semi-major axis $a$ and
orbital eccentricity $e$. The first star, with mass $M_1$ and
radius $R_1$, is rotating uniformly around an axis perpendicular to
the orbital plane in the sense of the orbital motion. The angular
velocity $\vec{\Omega}$ is assumed to be low so that the effects of
the Coriolis force and the centrifugal force can be neglected. The
second star, with mass $M_2$, is treated as a point mass.

We start from the right-handed orthogonal frame of reference $C_1
x^{\prime 1} x^{\prime 2} x^{\prime 3}$ introduced in Paper~I. The
origin of the frame of reference coincides with the mass centre $C_1$
of the uniformly rotating star and the $x^{\prime 1} x^{\prime
2}$-plane corresponds to the orbital plane of the binary. The
directions of the $x^{\prime 1}$- and the $x^{\prime 3}$-axis
coincide with the direction from the star's mass centre to the
periastron in the companion's relative orbit and to the direction
of the star's angular velocity $\vec{\Omega}$, respectively.

As in Paper~I, we pass on to an orthogonal frame of reference whose origin 
and $x^{\prime \prime 3}$-axis correspond to the origin and the 
$x^{\prime 3}$-axis of the frame of
reference $C_1 x^{\prime 1} x^{\prime 2} x^{\prime 3}$, but whose
$x^{\prime \prime 1}$-axis and $x^{\prime \prime 2}$-axis are
corotating with the star. We use this frame of reference to describe
the nonradial oscillations induced in the star by the orbiting
companion. With respect to the corotating frame of reference, we
introduce the spherical coordinates $\vec{r}=(r,\theta,\phi)$.  The
transformation formulae from the Cartesian coordinates $x^{\prime 1}$,
$x^{\prime 2}$, $x^{\prime 3}$ to the spherical coordinates $r$, $\theta$, $\phi$ are given by
\begin{equation}
\renewcommand{\arraystretch}{1.5}
\left.
\begin{array}{l c l}
x^{\prime 1} & = & r\, \sin \theta\, 
  \cos \left( \phi + \Omega\, t \right),  \nonumber \\
x^{\prime 2} & = & r\, \sin \theta\, 
  \sin \left( \phi + \Omega\, t \right),  \nonumber \\  
x^{\prime 3} & = & r\, \cos \theta. 
\end{array}\right\}  \label{trans1}
\end{equation}

In order to study the radial-velocity variations due to the tidal
motions of the mass elements located at the star's surface, it is
convenient to pass on to a frame of reference whose origin coincides
with the mass centre of the tidally distorted star and whose polar
axis is directed from the mass centre to the observer. We perform this transformation in two steps. 

First, we introduce 
an orthogonal frame of reference $C_1 \alpha^1 \alpha^2
\alpha^3$ whose origin and $\alpha^3$-axis coincide with the origin
and the $x^{\prime 3}$-axis of the frame of reference $C_1 x^{\prime
  1} x^{\prime 2} x^{\prime 3}$. The direction of the
$\alpha^1$-axis corresponds to the direction from the star's mass
centre to the ascending node in the relative orbit of the companion.  

Next, we pass on to an orthogonal frame of reference $C_1
\alpha^{\prime 1} \alpha^{\prime 2} \alpha^{\prime 3}$ with the same
origin and the same $\alpha^{\prime 1}$-axis as the frame of reference 
$C_1 \alpha^1 \alpha^2 \alpha^3$, but with its $\alpha^{\prime
3}$-axis directed from the star's mass centre to the observer. The
Cartesian coordinates $\alpha^{\prime 1}$, $\alpha^{\prime 2}$,
$\alpha^{\prime 3}$ are related to the
Cartesian coordinates $x^{\prime 1}$, $x^{\prime 2}$, $x^{\prime 3}$
by the transformation formulae
\begin{equation}
\left( \renewcommand{\arraystretch}{1.5}
\begin{array}{c}
\! \alpha^{\prime 1} \! \\
\! \alpha^{\prime 2} \! \\
\! \alpha^{\prime 3} \! 
\end{array} \right) = 
\left( \renewcommand{\arraystretch}{1.5}
\begin{array}{ccc}
\!\! \cos \omega & - \sin \omega & 0 \! \\
\!\! \sin \omega \cos i & \cos \omega \cos i & \sin i \! \\
\!\! - \sin \omega \sin i & - \cos \omega \sin i & \cos i \! 
\end{array} \right) \!
\left( \renewcommand{\arraystretch}{1.5}
\begin{array}{c}
\! x^{\prime 1} \!\! \\
\! x^{\prime 2} \!\! \\
\! x^{\prime 3} \!\! 
\end{array} \right).   \label{trans2}
\end{equation}
Here $\omega$ is the longitude of the periastron in the relative
orbit of the companion, and $i$ is the inclination of the orbital plane
with respect to the plane perpendicular to the line of sight 
\citep[see, e.g.,][]{Gre1985}. With
respect to the frame of reference $C_1 \alpha^{\prime 1}
\alpha^{\prime 2} \alpha^{\prime 3}$, we introduce the spherical 
coordinates $\vec{r}=(r^\prime,\theta^\prime,\phi^\prime)$. 

The geometry involved in the introduction of the various frames of
reference is illustrated in Fig.~\ref{frames}. 

\begin{figure}
\resizebox{\hsize}{!}{\includegraphics{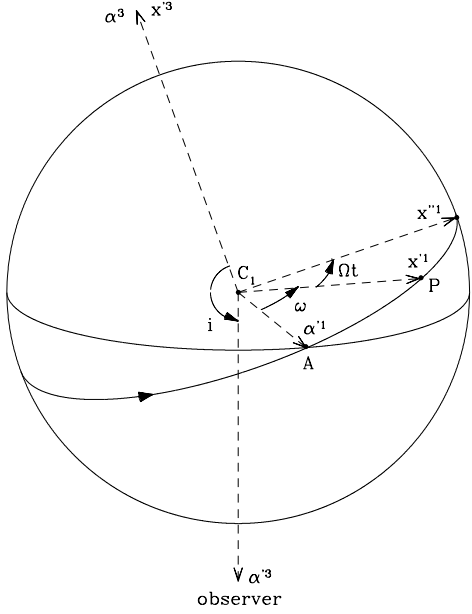}}
\caption{The geometry involved in the introduction of the various
frames of reference. The ascending node and the periastron of the
relative orbit of the companion are denoted by the points labeled 
A and P, respectively.} 
\label{frames}
\end{figure}

\section{The tide-generating potential}

\label{tidepot}

The tidal force exerted by the companion is derived from the 
tide-generating potential $\varepsilon_T\, W \left( \vec{r},t \right)$, 
where $\varepsilon_T$ is a small dimensionless parameter defined as 
\begin{equation}
\varepsilon_T = \left( {R_1 \over a} \right)^3\, {M_2 \over M_1}.
  \label{epsT}
\end{equation}
This parameter corresponds to the ratio of the tidal force to the 
gravity at the star's equator.

Following \citet{Pol1990}, we expand the tide-generating potential in 
terms of unnormalised spherical harmonics $Y_\ell^m(\theta,\phi)$ and 
in Fourier series in terms of multiples of the companion's mean motion
$n=2\pi/T_{\rm orb}$, where $T_{\rm orb}$
is the orbital period. The expansion can be written as
\begin{eqnarray}
\lefteqn{\varepsilon_T\, W \left( \vec{r},t \right) = 
  \varepsilon_T\, \sum_{\ell=2}^4 \sum_{m=-\ell}^\ell 
  \sum_{k=-\infty}^\infty W_{\ell,m,k} \left( \vec{r}\, \right)
  } \nonumber \\
& & \exp \left[ {\rm i}
  \left( \sigma_T\, t - k\, n\, \tau \right) \right],
  \label{pot}
\end{eqnarray}
where $\sigma_T = k\, n + m\, \Omega$ is the forcing angular frequency 
with respect to the corotating frame of reference, $\tau$ is a time of 
periastron passage, and 
\begin{equation}
W_{\ell,m,k} \left( \vec{r}\, \right) = - {{G\, M_1} \over R_1}\,
  c_{\ell,m,k}\, \left( {r \over R_1} \right)^\ell
  Y_\ell^m (\theta,\phi).  \label{pot2}
\end{equation}
Here $G$ is the Newtonian constant of gravitation 
and $c_{\ell,m,k}$ a Fourier coefficient determined as
\begin{eqnarray}
\lefteqn{c_{\ell,m,k} = \displaystyle
  {{(\ell-|m|)!} \over {(\ell+|m|)!}}\, P_\ell^{|m|}(0)
  \left({R_1\over a}\right)^{\ell-2}
  {1\over {\left({1 - e^2}\right)^{\ell - 1/2}}} } \nonumber \\
 & & {1\over \pi} \int_0^\pi (1 + e\, \cos v)^{\ell-1}\,
  \cos (k\, M + m\, v)\, dv. \label{pot:2}
\end{eqnarray}
In the latter expression, $P_\ell^m(x)$ is an associated Legendre
polynomial of the first kind, and $M$ and $v$ are respectively the
mean and the true anomaly of the companion in its relative orbit.  

The Fourier coefficients $c_{\ell,m,k}$ decrease with increasing
values of the multiple $k$ of the companion's mean motion. The
decrease is slower for higher orbital eccentricities, so that the
number of terms that has to be taken into account in
Expansion~(\ref{pot}) of the tide-generating potential increases with
increasing values of the orbital eccentricity. For $\ell=2$, the only
non-zero coefficients $c_{2,m,k}$ are those associated with
$m=-2,0,2$. A more elaborate discussion of these coefficients can be
found in \citet{W2000}.  

For realistic orbital periods, the forcing angular frequencies
$\sigma_T$ may be close to the eigenfrequencies of the star's free
oscillation modes $g^+$. The tidal action exerted by the
companion is then enhanced and the oscillation mode involved is 
resonantly excited with the forcing frequency of the  
dynamic tide (Paper~I). The effects of resonant excitations of modes in close binaries are
particularly important in the cases of resonances of dynamic tides
with $g^+$-modes of a lower radial order.

\section{Tidal displacement due to resonant dynamic tides}

We consider a single partial dynamic tide generated by the term
\begin{equation}
\varepsilon_T\, W_{\ell,m,k}\left(\vec{r}\,\right) \exp \left[
  {\rm i} \left( \sigma_T\, t - k\, n\, \tau \right) \right]
  \label{t1}
\end{equation}
in Expansion (\ref{pot}) of the tide-generating potential and assume
the forcing angular frequency $\sigma_T$ to be close to the
eigenfrequency $\sigma_{\ell,N}$ of a free isentropic oscillation mode
of radial order $N$ that is associated with the spherical harmonic
$Y_\ell^m(\theta,\phi)$.  
For the sake of simplification, we introduce the abbreviation $S =
(\ell,m,N)$ for the wave numbers $\ell$, $m$, and $N$ identifying the
oscillation mode involved in the resonance.  

Furthermore, let the relative frequency difference
\begin{equation}
\varepsilon = {{\sigma_{\ell,N} - \sigma_T}\over {\sigma_{\ell,N}}}  
  \label{eps}
\end{equation}
be of the order
of the ratio $\varepsilon_T$ of the tidal force to the gravity at the
star's equator. At the lowest-order of approximation, the
components of the resonant dynamic tide with respect to the local
coordinate basis $\partial/\partial r$, $\partial/\partial \theta$,
$\partial/\partial \phi$ are then given by
\begin{equation} 
\renewcommand{\arraystretch}{2.2} 
\left. \begin{array}{l c l} 
\lefteqn{ (\delta r)_T\left(\vec{r},t\right) = 
  {1 \over \varepsilon}\, {\varepsilon_T \over 2}\, 
  c_{\ell,m,k}\, Q_{\ell,N}\, \xi_{\ell,N}(r)} \nonumber \\
 & & P_\ell^{|m|}(\cos \theta)\, \exp \left[ {\rm i} 
  \left(m\,\phi + \sigma_T\,t - k\,n\,\tau \right) \right], 
  \nonumber \\ 
\lefteqn{ (\delta \theta)_T\left(\vec{r},t\right) = 
  {1 \over \varepsilon}\, {\varepsilon_T \over 2}\, 
  c_{\ell,m,k}\, Q_{\ell,N}\, {{\eta_{\ell,N}(r)}\over r^2} } 
  \nonumber \\
 & & \displaystyle {{\partial P_\ell^{|m|}(\cos \theta)} \over 
  {\partial \theta}}\, \exp \left[ {\rm i} 
  \left(m\,\phi + \sigma_T\,t - k\,n\,\tau \right) \right], 
  \nonumber \\ 
\lefteqn{ (\delta \phi)_T\left(\vec{r},t\right) =  
  {1 \over \varepsilon}\, {\varepsilon_T \over 2}\, 
  c_{\ell,m,k}\,  Q_{\ell,N}\, 
  {{\eta_{\ell,N}(r)} \over r^2} }  \nonumber \\
 & & \displaystyle {{{\rm i}\, m} \over {\sin^2 \theta}}\, 
  P_\ell^{|m|}(\cos \theta)\, \exp \left[ {\rm i} 
  \left(m\,\phi + \sigma_T\,t - k\,n\,\tau \right) \right]    
\end{array} \right\}  \label{res:1} 
\end{equation}
[Paper~I, Eqs.\ (57)].  
Here $\xi_{\ell,N}(r)$ and $\eta_{\ell,N}(r)$ are the radial parts of the 
radial and the transverse component of the Lagrangian displacement of the 
oscillation mode $S$ [Paper~I, Eqs.\ (43)], and $Q_{\ell,N}$ is related to the 
work done by the tidal force through the oscillation mode $S$ [Paper~I, 
Eq.\ (50)].

The components of the resonant dynamic tide with respect to the local
coordinate basis $\partial/\partial r^\prime$, $\partial/\partial
\theta^\prime$, $\partial/\partial \phi^\prime$ are obtained 
from Solutions (\ref{res:1}) by
application of the transformation formulae for contravariant vector
components. The transformation formula takes the form
\begin{equation}
\left( \renewcommand{\arraystretch}{1.5}
\begin{array}{c}
\! \left( \delta r^\prime \right)_T \! \\
\! \left( \delta \theta^\prime \right)_T \! \\
\! \left( \delta \phi^\prime \right)_T \! 
\end{array} \right) = 
\left( \renewcommand{\arraystretch}{1.9}
\begin{array}{ccc}
\!\! \displaystyle {{\partial r^\prime} \over {\partial r}}
 & \displaystyle {{\partial r^\prime} \over {\partial \theta}}
 &  \displaystyle{{\partial r^\prime} \over {\partial \phi}} \! \\
\!\! \displaystyle {{\partial \theta^\prime} \over {\partial r}}
 & \displaystyle {{\partial \theta^\prime} \over {\partial \theta}}
 & \displaystyle {{\partial \theta^\prime} \over {\partial \phi}} \! \\
\!\! \displaystyle {{\partial \phi^\prime} \over {\partial r}}
 & \displaystyle {{\partial \phi^\prime} \over {\partial \theta}}
 & \displaystyle {{\partial \phi^\prime} \over {\partial \phi}} \! 
\end{array} \right) \!
\left( \renewcommand{\arraystretch}{1.5}
\begin{array}{c}
\! \left( \delta r \right)_T \!\! \\
\! \left( \delta \theta \right)_T \!\! \\
\! \left( \delta \phi \right)_T \!\! 
\end{array} \right),   \label{trans3a}
\end{equation}
where  
\begin{equation} 
\renewcommand{\arraystretch}{1.9} 
\left. \begin{array}{l c l} 
\displaystyle {{\partial r^\prime} \over {\partial r}} & = & 1,  \\
\displaystyle {{\partial \theta^\prime} \over {\partial \theta}} & = & 
  \displaystyle {{\cos i - \sin i \sin \phi^\prime \cot \theta^\prime} 
  \over {D \left( \theta^\prime, \phi^\prime \right)}},  \\
\displaystyle {{\partial \theta^\prime} \over {\partial \phi}} & = & 
  \sin i \cos \phi^\prime,  \\
\displaystyle {{\partial \phi^\prime} \over {\partial \theta}} & = & 
  \displaystyle - {{\sin i \cos \phi^\prime} \over {\sin^2 \theta^\prime\,
  D \left( \theta^\prime, \phi^\prime \right)}},  \\
\displaystyle {{\partial \phi^\prime} \over {\partial \phi}} & = & 
  \cos i - \sin i \sin \phi^\prime \cot \theta^\prime, \hspace{0.5cm}
\end{array} \right\}   \label{trans3bb}
\end{equation} 
and the other partial derivatives are equal to zero. 
The function $D \left( \theta^\prime, \phi^\prime \right)$ is given by 
\begin{equation}
D = \left[ 
   \cos^2 \phi^\prime + \left( \cos i\, \sin \phi^\prime
   - \sin i\, \cot \theta^\prime \right)^2 \right]^{1/2}
   \!\!\!. \label{D}
\end{equation}

In addition, the spherical harmonics $Y_\ell^m(\theta,\phi)$ 
are transformed according to 
\begin{eqnarray}
\lefteqn{Y_\ell^m(\theta,\phi) = \exp \left[ - {\rm i}\, 
  m \left( \omega + \Omega\, t + \pi/2 \right) \right] } \nonumber \\
 & & \sum_{k^\prime=-\ell}^\ell  
   a_{\ell,m,k^\prime}(i)\, 
   Y_\ell^{k^\prime}(\theta^\prime,\phi^\prime), \label{trans4b}
\end{eqnarray}
where the coefficients $a_{\ell,m,k^\prime}(i)$ are functions of the orbital inclination defined as 
\begin{eqnarray}
\lefteqn{a_{\ell,m,k^\prime}(i) = \mu_m^{-1}\, \mu_{k^\prime}\,
 (\ell+m)!\, (\ell-k^\prime)! 
 \, \exp ({\rm i}\, k^\prime \pi/2) }  \nonumber \\
 & & \sum_j {{(-1)^{\ell-m-j} 
 \left(\cos{i \over 2}\right)^{2j+m+k^\prime}
   \left(\sin{i \over 2}\right)^{2\ell-2j-m-k^\prime}}
   \over {j!\, (\ell-m-j)!\, (\ell-k^\prime-j)!\, (m+k^\prime+j)!}}
   \label{sph:3}
\end{eqnarray}
\citep{Jef1965}. 
In this definition,  
the summation is performed over all values of $j$ satisfying the relations
\[
j \ge 0,\, j \ge -m-k^\prime,\, j \le \ell-m,\, j \le \ell-k^\prime,
\]  
and the factors $\mu_{k^\prime}$ are given by
\begin{equation}
\renewcommand{\arraystretch}{2.0}
\left.
\begin{array}{l c l}
\mu_{k^\prime} = 1 & \, {\rm for} & k^\prime \ge 0, \\
\mu_{k^\prime} = (-1)^{k^\prime}\, 
  \displaystyle {{(\ell+k^\prime)!}\over{(\ell-k^\prime)!}} 
  & \, {\rm for} & k^\prime < 0.
\end{array} \hspace{0.3cm} \right\}  \label{sph:6}
\end{equation}

At the lowest order of approximation,  
the components of the resonant dynamic tide with respect to the local
coordinate basis $\partial/\partial r^\prime$, $\partial/\partial
\theta^\prime$, $\partial/\partial \phi^\prime$ then take the form
\begin{equation}
\renewcommand{\arraystretch}{2.2}
\left.
\begin{array}{l c l}
\lefteqn{ (\delta r^\prime)_T\left(\vec{r}^\prime,t\right) =  
  {1 \over \varepsilon}\, {\varepsilon_T \over 2}\, 
  c_{\ell,m,k}\, Q_{\ell,N}\, \xi_{\ell,N}(r^\prime) 
  }  \nonumber \\
 & & \displaystyle 
  \sum_{k^\prime=-\ell}^\ell  
  a_{\ell,m,k^\prime}(i)\, P_\ell^{|k^\prime|}(\cos \theta^\prime)
  \nonumber \\
 & & \displaystyle \exp \left\{ {\rm i} \left[
  k^\prime\, \phi^\prime + k\, M   
  - m \left(\omega + {\pi \over 2} \right) \right]
  \right\}, \nonumber \\
\lefteqn{ (\delta \theta^\prime)_T\left(\vec{r}^\prime,t\right) = 
  {1 \over \varepsilon}\, {\varepsilon_T \over 2}\, c_{\ell,m,k}\,
  Q_{\ell,N}\, {{\eta_{\ell,N}(r^\prime)}\over r^{\prime 2}}
  } \nonumber \\
 & & \displaystyle  
  \sum_{k^\prime=-\ell}^\ell  
  a_{\ell,m,k^\prime}(i)\, {{\partial P_\ell^{|k^\prime|}
  (\cos \theta^\prime)}\over{\partial \theta^\prime}}
  \nonumber \\
 & & \displaystyle \exp \left\{ {\rm i} \left[
  k^\prime\, \phi^\prime + k\, M  
  - m \left(\omega + {\pi \over 2} \right) \right]
  \right\}, \nonumber \\
\lefteqn{ (\delta \phi^\prime)_T\left(\vec{r}^\prime,t\right) =   
  {1 \over \varepsilon}\, {\varepsilon_T \over 2}\, c_{\ell,m,k}\,
  Q_{\ell,N}\, {{\eta_{\ell,N}(r^\prime)} \over {
  r^{\prime 2}\, sin^2 \theta^\prime}}\, 
  }  \nonumber \\
 & & \displaystyle 
  \sum_{k^\prime=-\ell}^\ell  
  a_{\ell,m,k^\prime}(i)\, {\rm i}\, k^\prime\,
  P_\ell^{|k^\prime|}(\cos \theta^\prime)\,  \nonumber \\
 & & \displaystyle \exp \left\{ {\rm i} \left[
  k^\prime\, \phi^\prime  + k\, M   
  - m \left(\omega + {\pi \over 2} \right) \right]
  \right\}. 
\end{array} \hspace{0.8cm} \right\}  \label{vrad:4}
\end{equation}

\section{Radial-velocity variations associated with resonant 
dynamic tides}

The tidal motions of the mass elements located at the star's surface
contribute to the radial-velocity variations seen by the observer.  
Since the mass elements of the undistorted equilibrium star are assumed 
to be at rest with respect to the corotating frame of reference, 
the radial component of the velocity field associated with the tidal
motions is given by 
\begin{equation}
v_T = {{\partial \left(\delta \vec{r}\right)_T} \over 
  {\partial t}} \cdot \left( - \vec{e}_{\alpha^{\prime 3}} \right)
  = {\rm i}\, \sigma_T\, \left(\delta \vec{r}\right)_T 
  \cdot \left( - \vec{e}_{\alpha^{\prime 3}} \right),  \label{v:2}
\end{equation}
where 
\begin{equation}
\vec{e}_{\alpha^{\prime 3}} =  \cos \theta^\prime\, 
  {\partial \over {\partial r^\prime}} - 
  {{\sin \theta^\prime} \over r^\prime}\, 
  {\partial \over {\partial \theta^\prime}}  \label{v:3}
\end{equation}
is a unit vector in the direction of the $\alpha^{\prime 3}$-axis.

In the case of a resonance between a dynamic tide and a free oscillation mode, the tidally induced radial-velocity variations due to the mass elements located at the star's surface take the form 
\begin{eqnarray}
\lefteqn{v_{T;{\rm res}} \left(R_1, \theta^\prime, \phi^\prime; t \right) = -
  {1 \over \varepsilon}\, {\varepsilon_T \over 2}\, c_{\ell,m,k}\,
  Q_{\ell,N}\, {\rm i}\, \sigma_T }  \nonumber \\
 & & \sum_{k^\prime=-\ell}^\ell  
  a_{\ell,m,k^\prime}(i) \Bigg[ \cos \theta^\prime\, 
  \xi_{\ell,N} \left( R_1 \right)\, 
  P_\ell^{|k^\prime|}(\cos \theta^\prime) \nonumber \\
 & & \left. - \sin \theta^\prime\, 
  {{\eta_{\ell,N}\left( R_1 \right)} \over {R_1}}\, 
  {{\partial P_\ell^{|k^\prime|}
  (\cos \theta^\prime)}\over{\partial \theta^\prime}} \right] 
  \nonumber \\
 & & \displaystyle \exp \left\{ {\rm i} \left[
  k^\prime\, \phi^\prime + k\, M  
  - m \left(\omega + {\pi \over 2} \right)
  \right] \right\}.  \label{v:4}
\end{eqnarray}

The velocity variations seen by the observer are obtained by taking
the average of Eq.\ (\ref{v:4}) over the visible hemisphere of the
star. When  
one neglects the distortion of the stellar surface and the perturbation 
of the specific intensity due to the tidal motions of the mass elements, 
the average is determined as
\begin{eqnarray}
\lefteqn{V_{T;{\rm res}}(t) = } \nonumber \\
 & & {{\int_0^{2\pi} \int_0^{\pi/2} 
  h \left( \theta^\prime \right) 
  v_{T;{\rm res}}\! \left(R_1, \theta^\prime, \phi^\prime; t \right)
  \sin \theta^\prime \cos \theta^\prime 
  d\theta^\prime d\phi^\prime}
  \over {\int_0^{2\pi} \int_0^{\pi/2} h \left(\theta^\prime \right)
  \sin \theta^\prime \cos \theta^\prime d\theta^\prime d\phi^\prime}}, 
  \label{v:8}
\end{eqnarray}
where $h \left(\theta^\prime \right)$ is a limb-darkening law of
the form 
\begin{equation}
h \left(\theta^\prime \right) = 1 - u + u\, 
  \cos \theta^\prime, \;\;\;\;\; u \in [0,1]. \label{v:7}
\end{equation}
After performing the integration over the azimuthal angle
$\phi^\prime$, one obtains  
\begin{eqnarray}
\lefteqn{V_{T;{\rm res}}(t) = - {1 \over \varepsilon}\, 
  {\varepsilon_T \over 2}\, c_{\ell,m,k}\, {\rm i}\,
  \sigma_T\, a_{\ell,m,0}(i) }  \nonumber \\
 & & \left[ f_{\xi,\ell}(u)\, \xi_{\ell,N}\! \left(R_1\right)
  - f_{\eta,\ell}(u)\,
  {{\eta_{\ell,N}\! \left( R_1 \right)} \over {R_1}} 
  \right] Q_{\ell,N}  \nonumber \\
 & & \exp \left\{ {\rm i} \left[k\, M  
  - m \left(\omega + {\pi \over 2} \right) \right] \right\}.
  \label{v:9}
\end{eqnarray}
Here $f_{\xi,\ell}(u)$ and $f_{\eta,\ell}(u)$ are functions of the
limb-darkening coefficient $u$ defined as
\begin{eqnarray}
\lefteqn{f_{\xi,\ell}(u) = {6 \over {3-u}} }  \nonumber \\
 & & \int_0^{\pi/2} \! h \left(\theta^\prime \right)  
  P_\ell \left(\cos \theta^\prime \right) \sin \theta^\prime 
  \cos^2 \theta^\prime\, d\theta^\prime,  \label{v:10a} \\
\lefteqn{f_{\eta,\ell}(u) = {6 \over {3-u}}\ }  \nonumber \\
 & &  \int_0^{\pi/2} \!  \displaystyle 
  h \left(\theta^\prime \right)  
  {{\partial P_\ell \left(\cos \theta^\prime \right)} 
  \over {\partial \theta^\prime}}\, \sin^2 \theta^\prime\, 
  \cos \theta^\prime\, d\theta^\prime.  \label{v:10b}
\end{eqnarray}
The functions 
are described in more detail in Appendix~\ref{fxe}.

In addition to the partial dynamic tide with forcing angular frequency
$\sigma_T$, we also take into account the partial dynamic tide 
with forcing angular frequency $-\sigma_T$, which is also resonant. 
By the use of the symmetry properties $c_{\ell,-m,-k} = c_{\ell,m,k}$ and
$a_{\ell,-m,0} = a_{\ell,m,0}$, the global
solution for the variation of the star's radial velocity due to a
resonance of a dynamic tide with a free oscillation mode takes the form 
\begin{eqnarray}
V_{T;{\rm res}}(t) =  \varepsilon_T\, a_{\ell,m,0}(i)\, c_{\ell,m,k}(e)\,
   V_{T;{\rm res}}^\ast(t), \label{v:9b}
\end{eqnarray}
where
\begin{eqnarray}
\lefteqn{V_{T;{\rm res}}^\ast(t) = {1 \over \varepsilon}\, 
  \sigma_T\, 
  \left[ f_{\xi,\ell}(u)\, \xi_{\ell,N}\! \left(R_1\right)
  - f_{\eta,\ell}(u)\,
  {{\eta_{\ell,N}\! \left( R_1 \right)} \over {R_1}} 
  \right]  }  \nonumber \\
 & & Q_{\ell,N}\, \sin \left[k\, n\, (t-\tau)  
  - m \left(\omega + {\pi \over 2} \right) \right].
  \label{v:9bb}
\end{eqnarray}
In this expression, the products $Q_{\ell,N}\, \xi_{\ell,N}(R_1)$ and
$Q_{\ell,N}\, \eta_{\ell,N}(R_1)$ are independent of the normalisation
adopted for the determination of the free oscillation mode of degree $\ell$
and radial order $N$ that is involved in the resonance. They also
decrease rapidly with increasing values of $N$ so
that the influence of higher-order modes on the tidally induced
radial-velocity variations is substantially smaller than that of
lower-order modes \citep{W1997a,W2000}.

\section{Radial-velocity variations associated with non-resonant
dynamic tides} 

We again consider a single partial dynamic tide generated by a term of
the form  
\begin{equation}
\varepsilon_T\, W_{\ell,m,k}\left(\vec{r}\,\right) \exp \left[
  {\rm i} \left( \sigma_T\, t - k\, n\, \tau \right) \right]
  \label{t3}
\end{equation}
in Expansion (\ref{pot}) of the tide-generating potential, but now
assume the forcing angular frequency $\sigma_T$ to be far from any of
the eigenfrequencies associated with the star's free isentropic
oscillation modes. The components of the tidal displacement field
generated by this term can be expressed as 
\begin{equation}
\renewcommand{\arraystretch}{2.0}
\left.
\begin{array}{l c l}
\lefteqn{(\delta r)_T \left(\vec{r},t\right) = 
  \varepsilon_T\, c_{\ell,m,k}\, \xi_T(r) }  \nonumber \\
 & &  P_\ell^{|m|}(\cos \theta)\, \exp \left[{\rm i} 
  \left( m\,\phi + \sigma_T\,t - k\,n\,\tau \right) \right],
  \nonumber \\
\lefteqn{(\delta \theta)_T \left(\vec{r},t\right) =
  \varepsilon_T\, c_{\ell,m,k}\, {{\eta_T(r)} \over r^2} }
  \nonumber \\
 & & \displaystyle  
  {{\partial P_\ell^{|m|}(\cos \theta)} \over {\partial \theta}}\, 
  \exp \left[{\rm i} \left( m\,\phi + \sigma_T\,t - k\,n\,\tau \right)
  \right], \nonumber \\
\lefteqn{(\delta \phi)_T \left(\vec{r},t\right) = 
  \varepsilon_T\, c_{\ell,m,k}\, {{\eta_T(r)} \over r^2}\, 
  {{{\rm i}\,m} \over {\sin^2 \theta}}}  \nonumber \\
 & & \displaystyle  P_\ell^{|m|}(\cos \theta)\,  
  \exp \left[{\rm i} \left( m\,\phi + \sigma_T\,t - k\,n\,\tau \right)
  \right], \nonumber
\end{array} \hspace{0.2cm} \right\}  \label{vrad:26}
\end{equation}
where $\xi_T(r)$ and $\eta_T(r)$ are solutions of the homogeneous 
fourth-order system of linear differential equations
\begin{eqnarray}
\lefteqn{ {{d\!\left( r^2\, \xi_T \right)}\over {dr}} = {g\over c^2}\,
  r^2\, \xi_T + \!\left[\ell (\ell+1) - \sigma_T^2\, {r^2\over c^2} 
  \right]\!
  \eta_T + {r^2\over c^2}\, \Psi_T, } \label{vrad:21} \\
\lefteqn{ {{d\eta_T}\over {dr}} = 
  \left( 1 - {N_b^2\over \sigma_T^2} \right) \xi_T
  + {N_b^2\over g}\, \eta_T - {1\over \sigma_T^2}\, {N_b^2\over g}\, \Psi_T, }
   \label{vrad:22} \\
\lefteqn{ {1\over r^2}\, {d\over{dr}} \left( r^2\, {{d\Psi_T}\over {dr}}
   \right) - {{\ell(\ell+1)}\over r^2}\, \Psi_T } \nonumber \\
  & = & 4\, \pi\, G\, \rho \left[ {N_b^2\over g}\, \xi_T + {1\over c^2}
   \left( \sigma_T^2\, \eta_T - \Psi_T \right) \right]. 
     \label{vrad:23}
\end{eqnarray}
Here $\Psi_T(r)$ is the radial part of the total perturbation of the 
gravitational potential \citep[for a definition see,
e.g.,][]{Pol1990}, $g$ the local  
gravity, $c^2$ the square of the isentropic sound velocity, and $N_b^2$ the 
square of the Brunt-V\"{a}is\"{a}l\"{a} frequency. 
The functions $\xi_T(r)$, $\eta_T(r)$, and $\Psi_T(r)$ depend on
the values of the azimuthal number $m$ and on the Fourier index $k$ through 
the forcing angular frequency $\sigma_T$. 

The solutions of Eqs.\ (\ref{vrad:21}) -- (\ref{vrad:23}) must 
satisfy boundary conditions at the star's centre and at the star's surface.  
At $r=0$, the radial component of the tidal displacement must remain finite. 
At $r=R_1$, the Lagrangian perturbation of the pressure must vanish, and the 
continuity of the gravitational potential and its gradient requires that
\begin{eqnarray}
\lefteqn{ \left({ {d\Psi_T}\over {dr} }\right)_{R_1} + 
  {{\ell+1} \over R_1}\, \left(\Psi_T \right)_{R_1} 
  + \left( 4\, \pi\, G\, \rho\, \xi_T \right)_{R_1}  } \nonumber \\ 
 & & = - (2\, \ell + 1)\, {{G\, M_1} \over R_1^2}.  \label{vrad:24}
\end{eqnarray}

By comparison of Eqs.\ (\ref{vrad:26}) with Eqs.\ (\ref{res:1}), it
follows that the contribution of a non-resonant dynamic tide to the
tidally induced radial-velocity variations takes the form 
\begin{eqnarray}
V_{T;\ell,m,k}(t) = \varepsilon_T\, 
  a_{\ell,m,0}(i)\, c_{\ell,m,k}(e)\, V_{T;\ell,m,k}^\ast(t)
  \label{vrad:31}
\end{eqnarray}
with
\begin{eqnarray}
\lefteqn{ V_{T;\ell,m,k}^\ast(t) =  }  \nonumber \\
 & & 2\, \sigma_T\, \left[ f_{\xi,\ell}(u)\, \xi_{T;\ell,m,k}\!   
  \left(R_1\right) - f_{\eta,\ell}(u)\, {{\eta_{T;\ell,m,k}\! 
  \left( R_1 \right)} \over {R_1}} \right] \nonumber \\
 & &  \sin \left[k\, n\, (t-\tau) - 
  m \left(\omega + {\pi \over 2} \right) \right].
  \label{vrad:31bb}
\end{eqnarray}

\section{Total variation of a star's radial velocity due to its tidal
 response} 

In accordance with Expansion (\ref{pot}) of the tide-generating potential, 
the total variation of the star's radial velocity due to the   
tidal motions of the mass elements located at its surface, is obtained
by adding the contributions stemming from the various resonant and
non-resonant partial dynamic tides.  

When one restricts the contributions to those resulting from the
tides associated with the second-degree spherical harmonics, the total
variation of the star's radial velocity due to its tidal response can
be written as
\begin{eqnarray}
\lefteqn{V_{T;{\rm rad}}(t) = \varepsilon_T  \sum_{k=1}^\infty 
  \Big\{ a_{2,0,0}(i)\, c_{2,0,k}(e)\, V_{T;2,0,k}^\ast(t) }
   \label{vrad:34} \\
 & & + a_{2,2,0}(i)\! \left[ c_{2,-2,k}(e) V_{T;2,-2,k}^\ast(t)
   + c_{2,2,k}(e) V_{T;2,2,k}^\ast(t) \right] \!\!
    \Big\}, \nonumber 
\end{eqnarray} 
where
\begin{equation}
\renewcommand{\arraystretch}{2.0}
\left.
\begin{array}{l c l}
\lefteqn{a_{2,0,0}(i) = \displaystyle 
  {1 \over 2} \left( 3\, \cos^2 i -1 \right), } \nonumber \\
\lefteqn{a_{2,2,0}(i) = a_{2,-2,0}(i) = 3\, \sin^2 i.} \nonumber 
\end{array} \hspace{5cm} \right\}  \label{sph:7}
\end{equation}
In the case of a resonance of a dynamic tide characterised by the
azimuthal number $m$ and the Fourier index $k$ with a free oscillation
mode of radial order $N$, the factor $V_{T;2,m,k}^\ast(t)$ must be
replaced by the factor $V_{T;{\rm res}}^\ast(t)$.  

The contributions of the various partial dynamic tides to the observed
tidally induced radial-velocity variations depend on the orbital
inclination and on 
the orbital eccentricity through the coefficients $a_{2,m,0}(i)$ and
$c_{2,m,k}(e)$, respectively. 
The variation of the product of the coefficients $a_{2,m,0}(i)$ and
$c_{2,m,k}(e)$ is displayed in Fig.~\ref{ca2mk} as a function of $k$,
for the orbital inclinations $i=5^\circ$ and $i=85^\circ$, and the
orbital eccentricities $e=0.3$ and $e=0.5$. The product  
decreases with increasing values of $k$ due to the rapid decrease of the
Fourier coefficients $c_{2,m,k}(e)$ for higher-order harmonics in
Expansion (\ref{pot}) of the tide-generating potential. 

\begin{figure*}
\resizebox{\hsize}{!}{\rotatebox{270}{\includegraphics{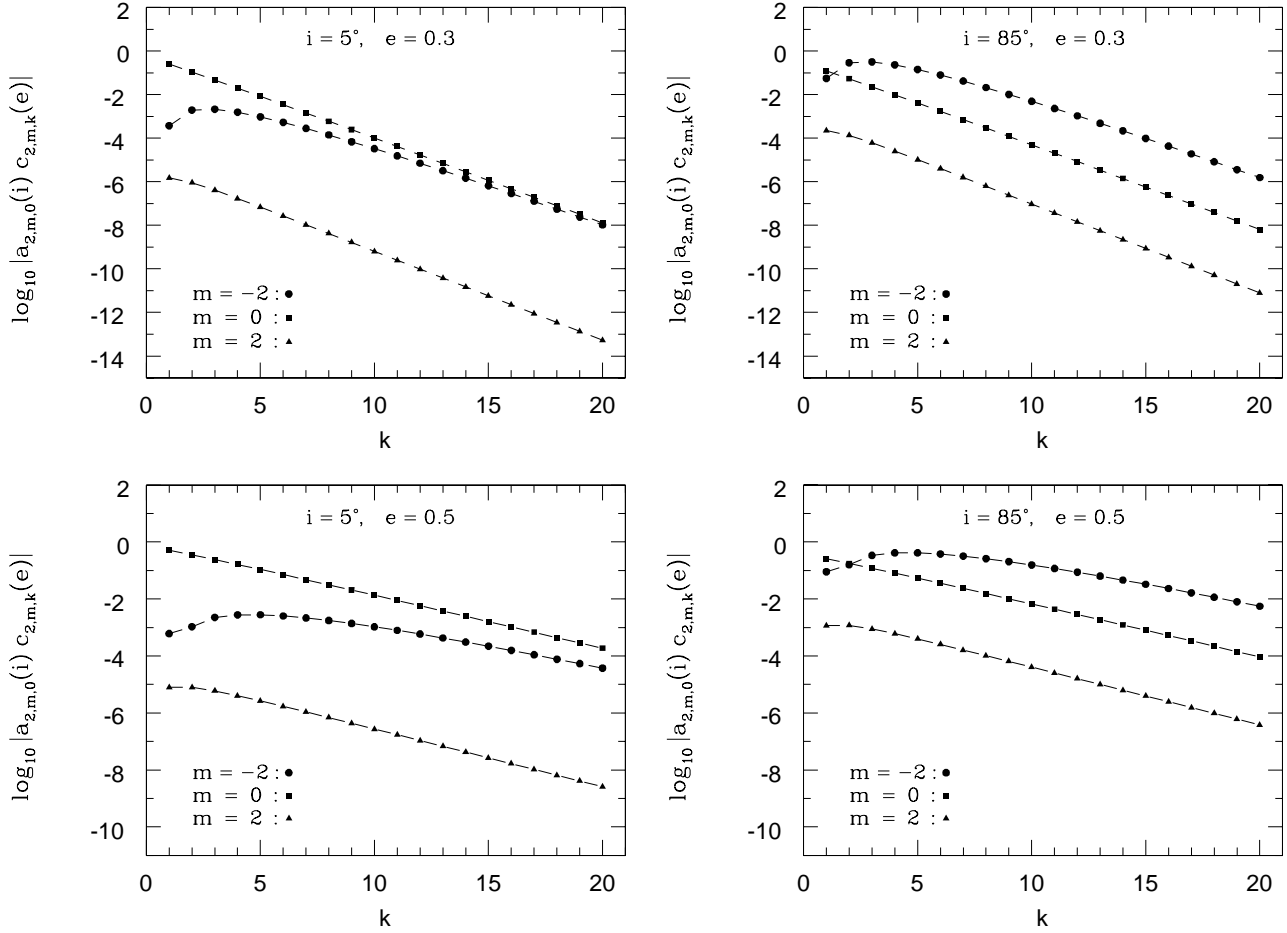}}}
\caption{Logarithmic representation of the absolute value of the
  products $a_{2,m,0}(i)\, c_{2,m,k}(e)$ for the orbital inclinations
  $i=5^\circ$ (left-hand panels) and $i=85^\circ$ (right-hand panels),
  and the orbital eccentricities $e=0.3$ (top panels) and $e=0.5$
  (bottom panels).}   
\label{ca2mk}
\end{figure*}

In the case of the lower orbital inclination $i=5^\circ$, the products
of the coefficients $a_{2,m,0}(i)$ and $c_{2,m,k}(e)$ are largest in
absolute value for the coefficients associated with the azimuthal
number $m=0$. The prevalence is more pronounced for the eccentricity
$e=0.5$ than for the eccentricity $e=0.3$ due to the increase of
the coefficients $c_{2,0,k}(e)$ with increasing values of the orbital
eccentricity. For larger values of the orbital inclination, the
products of the coefficients $a_{2,m,0}(i)$ and $c_{2,m,k}(i)$
associated with  $m=-2$ rapidly become the
dominant products. In the case of the orbital inclination
$i=85^\circ$, they are approximately two orders of magnitude larger in
absolute value than the products associated with the azimuthal number
$m=0$. The products of the coefficients 
associated with $m=2$ 
are generally two or more orders of magnitude smaller 
than the products associated with the azimuthal numbers $m=-2$ and $m=0$.

\section{Tidally induced radial-velocity variations in a $5\,M_\odot$
  zero-age main sequence star} 

We have applied the expressions derived in the previous sections to a
binary consisting of a $5\,M_\odot$ zero-age main sequence star and a
$1.4\, M_\odot$ companion which we approximate by a point mass.
We considered orbital periods ranging 
from 2 to 6 days and the orbital eccentricities $e=0.3$ and
$e=0.5$. The rotation of the star is assumed to be synchronised with
the orbital motion of the companion in the periastron of its relative
orbit. In addition, we have set both the longitude of the periastron
$\omega$ and the time of periastron passage $\tau$ equal to zero. For
the limb-darkening coefficient we adopted the value $u=0.36$, which is
appropriate for main-sequence B-type stars.

The observed amplitude of the tidally induced radial-velocity
variations is displayed in Figs.~\ref{ampl1} and \ref{ampl2} as a
function of the orbital period. The orbital inclination varies from
$i=0^\circ$ in the upper panels of the figures to $i=90^\circ$ in the
lower panels. The numerous peaks appearing at shorter orbital periods
correspond to resonances of dynamic tides with free oscillation
modes. Bearing in mind that, in the perturbation theory used in
Paper~I, the relative frequency
difference $\varepsilon$ is assumed to be of the order of
$\varepsilon_T$, the calculations near the resonances are restricted
to values of $\varepsilon$ larger than or equal to $0.1\,
\varepsilon_T$ in absolute value. 

\begin{figure}
\resizebox{8.5cm}{!}{\includegraphics{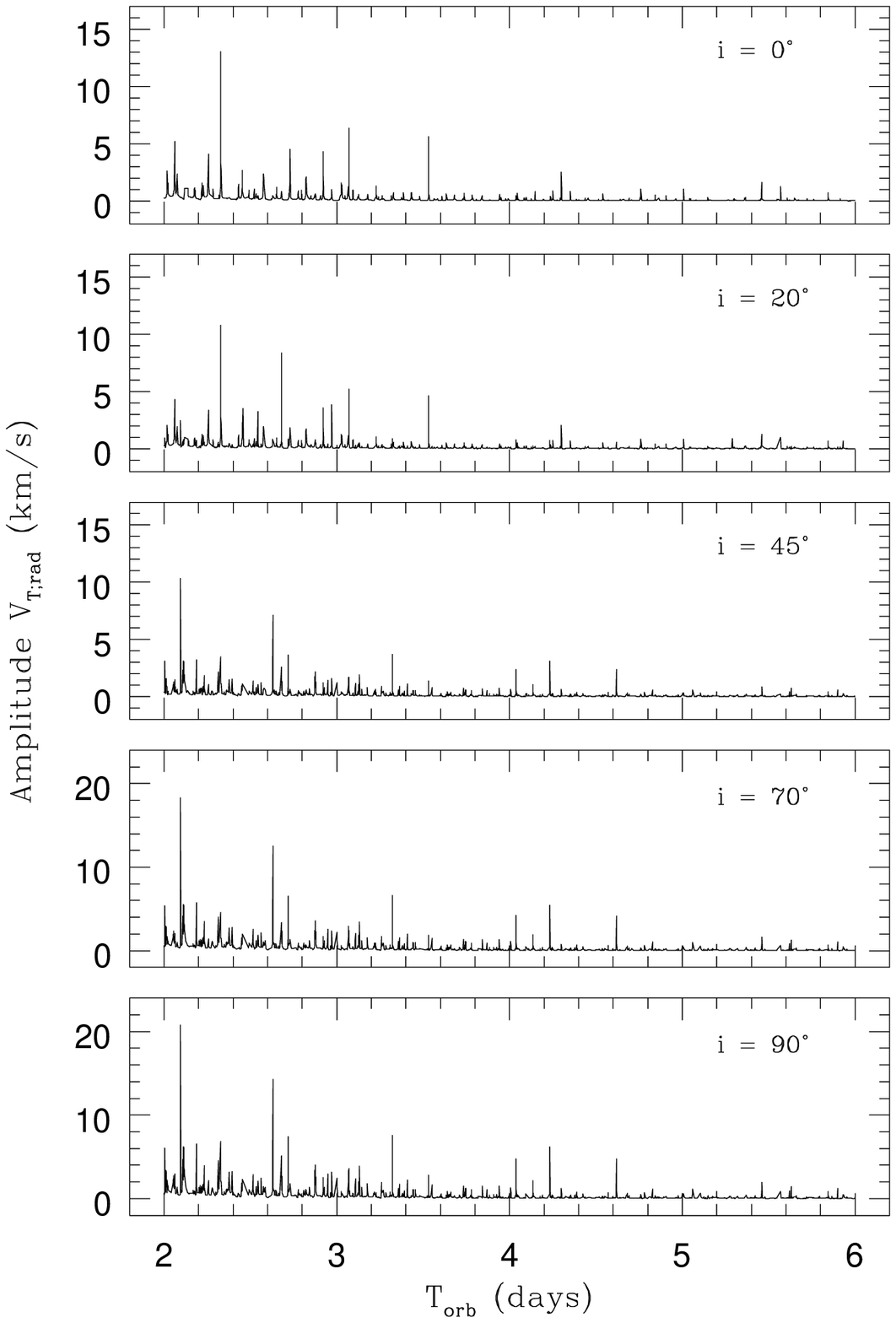}}
\caption{The observed amplitude of the tidally induced radial-velocity
  variations as a function of the orbital period for the $5\,M_\odot$
  ZAMS stellar model and the orbital eccentricity $e=0.3$.}   
\label{ampl1}
\end{figure}

\begin{figure}
\resizebox{8.5cm}{!}{\includegraphics{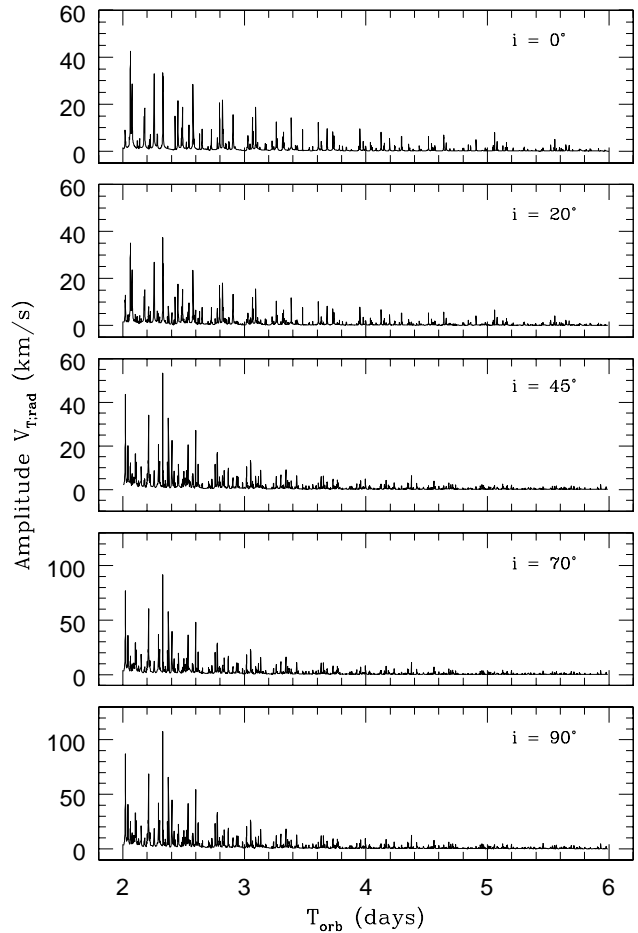}}
\caption{The observed amplitude of the tidally induced radial-velocity
  variations as a function of the orbital period for the $5\,M_\odot$
  ZAMS stellar model and the orbital eccentricity $e=0.5$.}   
\label{ampl2}
\end{figure}

When the orbital plane is observed face-on, i.e.\ when $i=0^\circ$, the
tidally induced radial-velocity variations are determined solely by
the dynamic tides associated with the azimuthal number $m=0$. For
higher values of the orbital inclination, additional contributions
arise from the dynamic tides associated with the azimuthal numbers
$m=\pm 2$, although the contributions associated with $m=2$ are
usually negligible. The amplitude of the tidally induced
radial-velocity variations is largest when the orbital plane is seen
edge-on, i.e.\ when $i = 90^\circ$. The resonances are then almost
exclusively due to the dynamic tides associated with the azimuthal
number $m=-2$.

For close resonances, the amplitudes of the tidally induced radial-velocity
variations can be quite large so that they are certainly detectable in
observations. They increase with increasing values of the orbital eccentricity
due to the smaller periastron distances associated with higher orbital
eccentricities.  The number of resonances is also larger
for the orbital eccentricity $e=0.5$ than for the orbital eccentricity
$e=0.3$ due to the larger number of terms that has to be taken into
account in Expansion~(\ref{pot}) of the tide-generating potential.
For some cases our results point towards very high amplitudes
which are above the typical sound speed in the atmosphere of the star. This
would imply shock waves to occur, but our linear theory breaks down at these
high velocities. The study of the nonlinear effects is currently beyond the
scope of our investigation and the applications are limited to 
radial-velocity variations that can be well described by the linear theory.

A detailed representation of the effects of resonances in the range of
orbital periods from 3.3 to 3.7 days is given in Fig.~\ref{detail} for
the orbital eccentricity $e=0.5$ and the orbital inclinations
$i=20^\circ$ and $i=70^\circ$. The strongest resonances observed give
rise to radial-velocity variations with amplitudes well above 
$5\,{\rm km\, s^{-1}}$. Several orbital periods lead to
resonances with two oscillation modes simultaneously. The peak in the
amplitude near $T_{\rm orb}=3.37$ days, for instance, is caused by
simultaneous resonances with the oscillation modes $g_{12}^+$ and
$g_{16}^+$. 

\begin{figure}
\resizebox{\hsize}{!}{\includegraphics{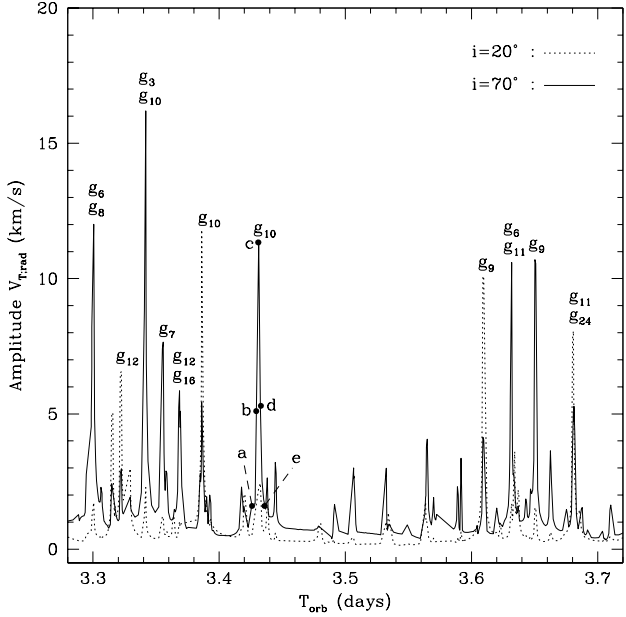}}
\caption{Detailed representation of the observed amplitude of the tidally
  induced radial-velocity variations of the $5\,M_\odot$ ZAMS stellar
  model for the orbital eccentricity $e=0.5$, and the orbital inclinations
  $i=20^\circ$ (dotted line) and $i=70^\circ$ (solid~line).} 
\label{detail}
\end{figure}

In Fig.~\ref{rad1}, the tidally induced radial-velocity variations
corresponding to the points labeled (a) -- (e) 
are shown as a function of the orbital phase. The central panel shows
that for an exact resonance -- point (c), the radial-velocity
variations are independent of the orbital phase. The reason is that
the tidally induced radial-velocity variations given by Expansion
(\ref{vrad:34}) are dominated by the term corresponding to the
resonant dynamic tide. Outside the resonance -- cases (a) and (e), the
radial-velocity variations are markedly less regular functions of the
orbital phase since they are now the result of a superposition of
purely non-resonant dynamic tides. 

\begin{figure}
\resizebox{\hsize}{!}{\includegraphics{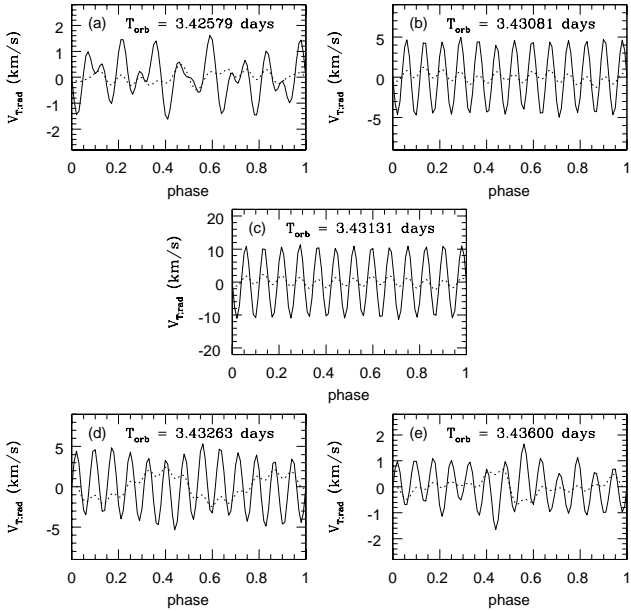}}
\caption{Observed tidally induced radial-velocity variations of the
  $5\,M_\odot$ ZAMS stellar model for orbital periods near 
  3.43 days, the orbital eccentricity $e=0.5$, and the orbital 
  inclinations $i=20^\circ$ (dashed line) and $i=70^\circ$ (solid line).} 
\label{rad1}
\end{figure}

\section{Comparison with observations: HD\,177863}

The bright slowly pulsating B star (SPB) HD\,177863 was discovered to
be a radial-velocity variable by \citet{Eggen1977}, while
\citet{WR1985} detected intrinsic photometric variability in this
object. Later on, \citet{W1991} classified the star as an SPB.
\citet{DeCat2000} showed this star to be a single-lined spectroscopic
binary and determined the orbital parameters of the system. They found
an orbital period $T_{\rm orb}= 11.9154 \pm 0.0009$ days, an 
eccentricity $e=0.60$, a longitude of the periastron
$\varpi=182^\circ$, and a time of periastron passage
$\tau=2450155.78\,{\rm JD}$.
The large eccentricity implies that the system is either still very
young or that it has undergone some episodes of mass transfer. We find
no indication of the occurrence of such episodes in the literature,
while the relatively young age is confirmed by the study of
\citet{West1985}, who finds an age of 64 million years. In what
follows, we therefore assume that the system consists of two
main-sequence stars.  From the spectroscopic analysis and the mass
function,
one can furthermore derive that the orbital inclination is between
35~and 90~degrees and that the companion has a mass smaller than
$2\,M_\odot$.  We also note that De Cat \& Aerts (in preparation) have
found the primary to rotate supersynchronously with a projected
equatorial rotation velocity between $45$ and $60\, {\rm km\,
s^{-1}}$.

\citet{DeCat2001} also performed a detailed frequency study on both
extensive spectroscopic and photometric time series and found two
intrinsic frequencies for the star: 0.84059\,c.d$^{-1}$ and
0.10108\,c.d$^{-1}$.  The first one of these differs less than
0.001\,c.d$^{-1}$ from 10 times the orbital frequency. This
observation, together with the binary configuration, led
\citet{DeCat2001} to conclude that one may be dealing with a
resonantly excited mode.

In order to investigate if the suggestion by \citet{DeCat2001} is
supported by the theory developed in our paper, we determined the
tidal response of a $3.5\, M_\odot$ stellar model for a fixed orbital
period of 11.9154 days as a function of the rotational angular
velocity $\Omega$. The model has an age of 64~million years,
corresponding to a central hydrogen abundance $X_c=0.6$ and a radius
$R_1=2.48\,R_\odot$. The limb-darkening coefficient is assumed to take
the value $u=0.36$. 

The resulting amplitudes of the tidally induced radial-velocity
variations are shown in Fig.~\ref{hd177863} for the orbital
inclinations $i=90^\circ$, $i=55^\circ$, and $i=35^\circ$. 
Since the expressions established in our theory are
derived in the linear approximation, we  restrict ourselves to the
presentation of radial-velocity variations with amplitudes smaller
than the sound speed in the atmosphere of the star. The range of
rotational angular velocities considered for each inclination is
determined from the range of projected equatorial rotation
velocities. The companion mass for each inclination results from the
mass function of the binary. 

\begin{figure*}
\resizebox{6cm}{!}{\includegraphics{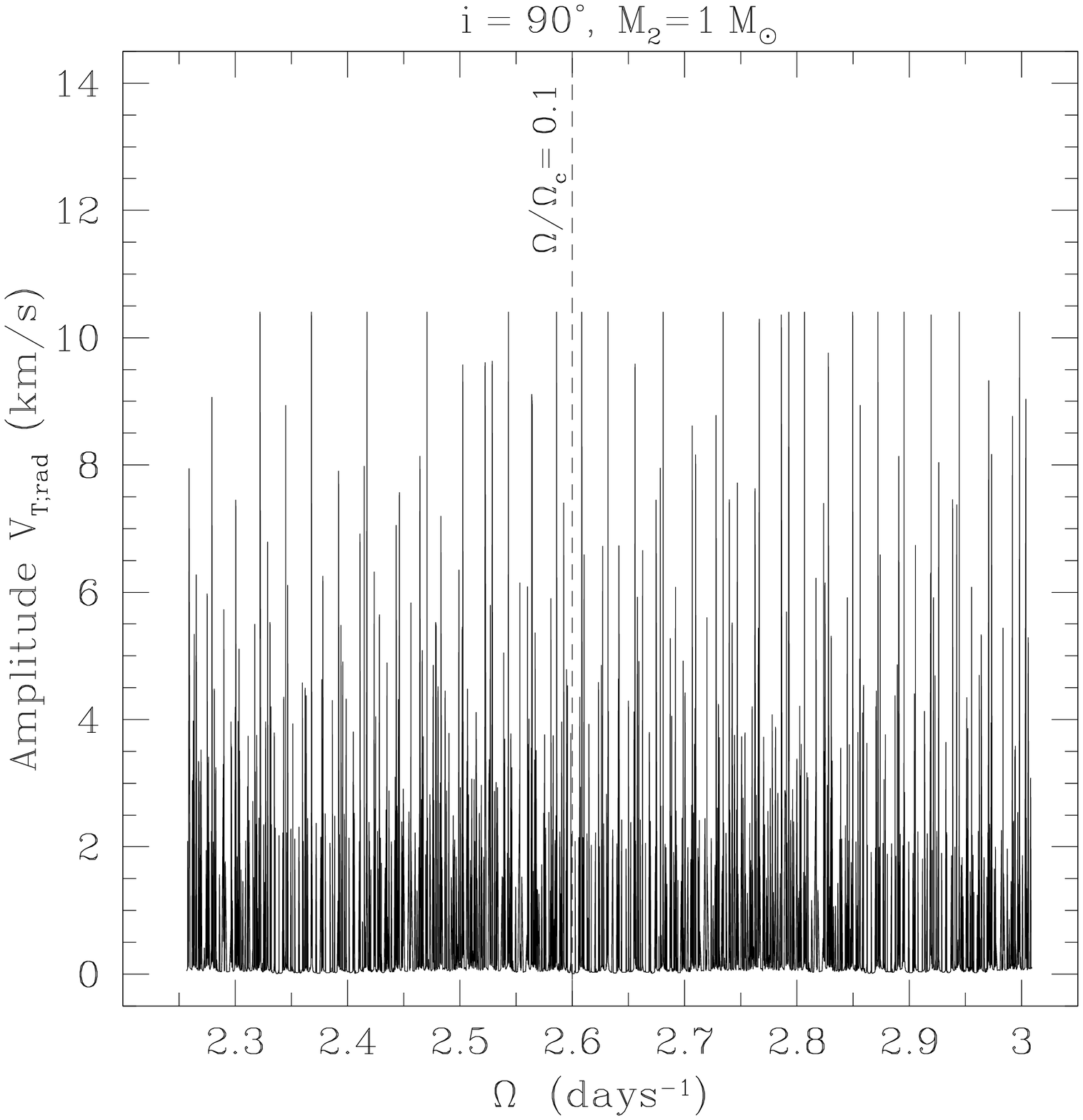}}
\resizebox{6cm}{!}{\includegraphics{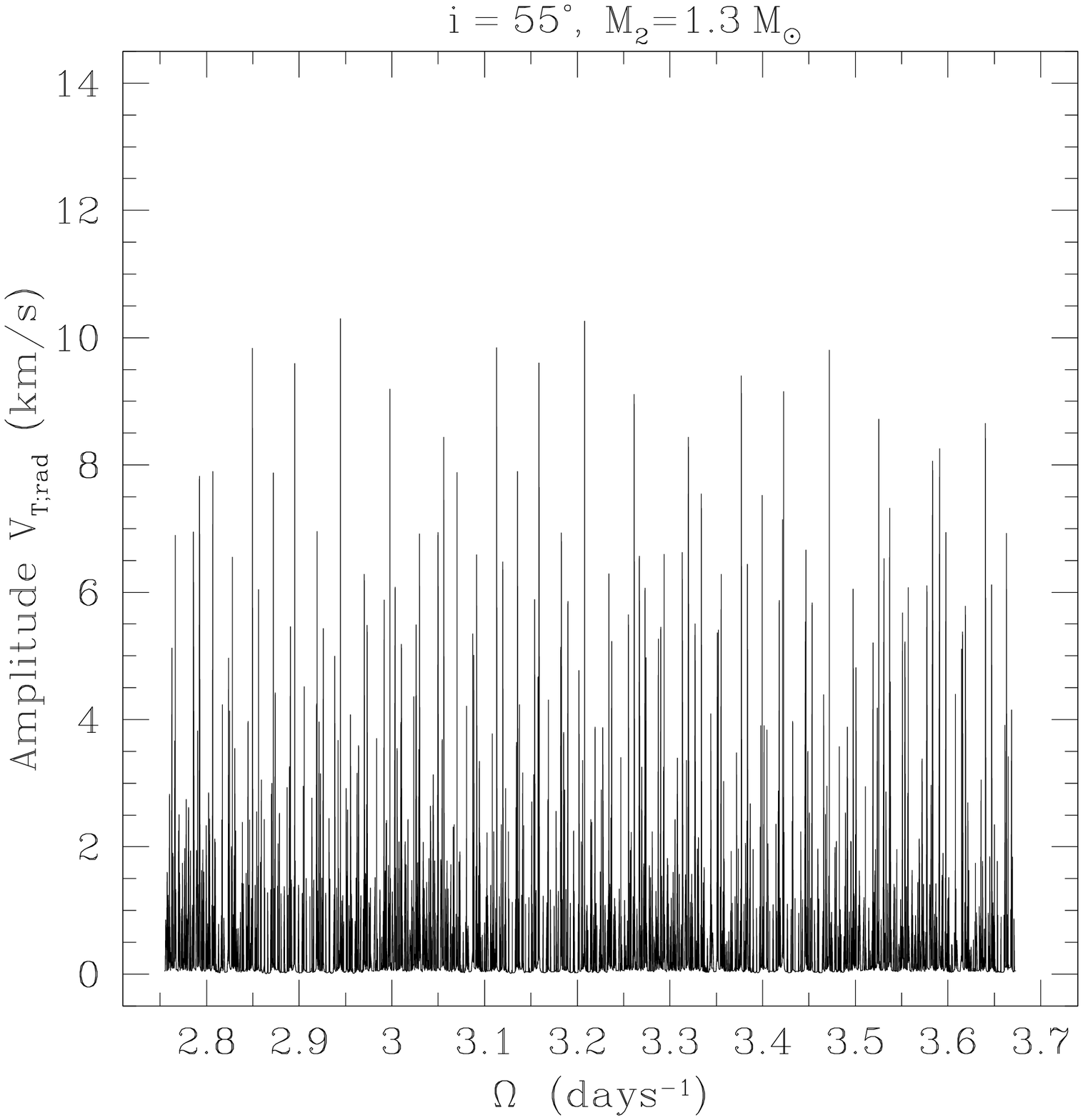}}
\resizebox{6cm}{!}{\includegraphics{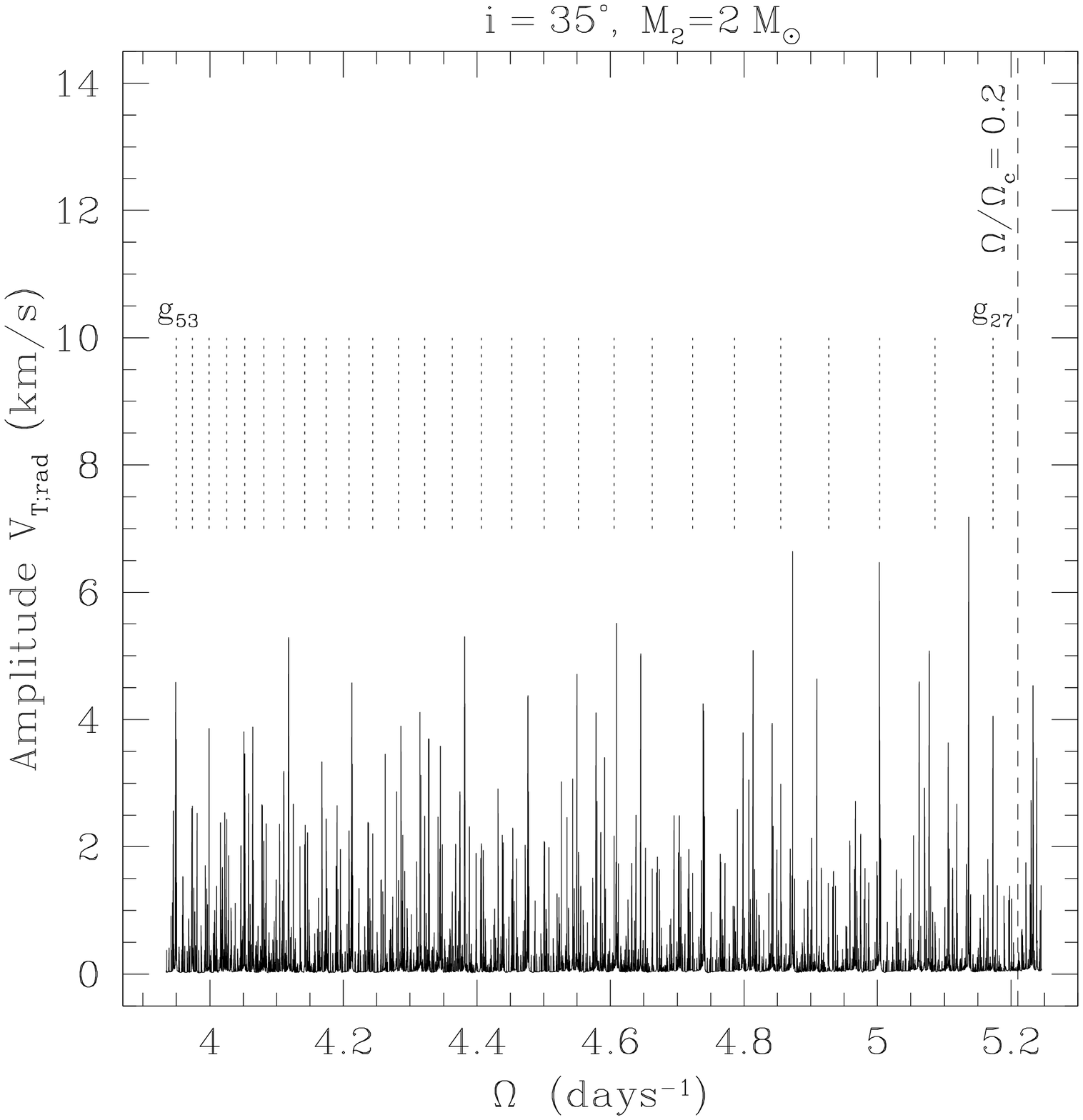}}
\caption{Observed amplitude of the tidally induced radial-velocity
  variations in HD\,177863, for the orbital inclinations $i=90^\circ$
  (left), $i=55^\circ$ (middle), and $i=35^\circ$ (right). The dashed
  lines in the left- and right-hand panels indicate the
  rotational angular velocities corresponding to 10\% and 20\% of the
  critical break-up angular velocity $\Omega_c$. The dotted lines in
  the right-hand panel indicate resonances caused by the partial
  dynamic tide associated with $m=-2$ and $k=10$. } \label{hd177863}  
\end{figure*}

A large number of resonances is found for all three orbital
inclinations, among which a whole range are caused by partial dynamic
tides associated with the Fourier index $k=10$ in
Expansion~(\ref{pot}) of the tide-generating potential. The azimuthal
number of the resonant dynamic tide takes the value $m=-2$ in all
these cases,  which is compatible with a preliminary mode
identification performed by \citet{DeCat2001}. While the photometric
data did not allow him to identify the mode, he found convincing
evidence from the variations of the moments of the line profiles
\citep[for a definition see, e.g.,][]{Aerts1992} to conclude that the
observed mode is most likely a sectoral $\ell=2$ mode. We note
that any resonances caused by partial dynamic tides associated with
the azimuthal number $m=0$ would have led to large amplitudes of the
tidally induced radial velocity variations for all values of the
rotational angular velocity. 

In the case of the orbital inclination $i=35^\circ$, we indicated the
positions of the resonances associated with $m=-2$ and $k=10$ by
dotted vertical lines in Fig.~\ref{hd177863}. We find a possible
resonant excitation of 
sectoral $\ell=2$ modes ranging from the mode $g_{27}^+$ up to the
mode $g_{53}^+$. For the orbital inclinations $i=55^\circ$ and
$i=90^\circ$, the radial orders of the modes excited by the partial
dynamic tide associated with $m=-2$ and $k=10$ are even higher.  

Upon close inspection, the pattern of resonances in 
Fig.~\ref{hd177863} can be seen to repeat itself 
at equidistantly spaced values of the rotational angular
velocity $\Omega$. The reason is easily identified when the forcing
angular frequency $\sigma_T$ is rewritten as  
\begin{equation}
k\, n + m\, \Omega = \left( k+1 \right) n 
  + m \left( \Omega - {n \over m} \right).  \label{pat}
\end{equation}
It follows that if the partial dynamic tide associated with the
azimuthal number $m$ and the Fourier index $k$ is resonant at the
rotational angular velocity $\Omega$, the partial dynamic tide
associated with the azimuthal number $m$ and the Fourier index $k+1$
will be resonant at the rotational angular velocity $\Omega-n/m$.

We conclude that due to the large uncertainty in the rotational angular
velocity and the orbital inclination of HD\,177863, a definitive mode
identification is not yet possible at this time. In addition, the
amplitude of the tidally induced radial-velocity variations can not be
determined in the isentropic approximation so that the inclusion of
nonadiabatic effects in our treatment becomes desirable. For higher
rotation rates, more resonances can also be expected due to the
rotational splitting of the stellar eigenfrequencies by the Coriolis
force. 
We plan to explore the influence of the nonadiabatic effects and the
Coriolis force in more detail in subsequent investigations. 
In particular, we will investigate the behaviour of the predicted 
observable quantities (photometric variations and line-profile variations)
resulting from the candidate resonant modes that we have found in this
work with the goal to confront them with the data. Such an iterative
procedure can perhaps allow us to come to a definite mode identification
and detailed modelling with the goal to constrain the internal 
structure parameters of the primary.

\section{Concluding remarks}

In this paper, we derived semi-analytical expressions for the tidally
induced radial-velocity variations in a uniformly rotating component
of a close binary. We neglect the effects of the Coriolis force and
the centrifugal force and treat both the free and the forced
oscillations of the component as linear, isentropic perturbations of a
spherically symmetric star. We take into account the possibility of
resonances between dynamic tides and free oscillation modes by means
of the perturbation theory developed by \citet{SWV1998}.

The amplitudes of the tidally induced radial-velocity variations
depend on the orbital eccentricity and on the inclination of the
orbital plane with respect to the plane perpendicular to the line of
sight. They increase with increasing values of the orbital
eccentricity and are largest when the orbital plane is seen edge-on.
The amplitude is most sensitive to the values of the 
orbital inclination when $20^\circ \la i \la 70^\circ$. A similar
conclusion was reached by \citet{Kum1995} in an investigation on the
stellar luminosity variations associated with tidally excited oscillation
modes.

From the application to a $5\,M_\odot$ zero-age main sequence star, it
follows that the amplitude of the tidally induced radial-velocity
variations seen by an observer is small, except when resonances occur
between dynamic tides and free oscillations modes. The resonances
enhance the tidal motions of the mass elements and lead to
radial-velocity variations with amplitudes that are up to an order of
magnitude larger than those observed outside resonances. Some orbital
periods are seen to give rise to simultaneous resonances with two
oscillation modes.

The shape of the tidally induced radial-velocity curves changes
markedly with the proximity of a dynamic tide to a resonance with a
free oscillation mode. For close resonances, the tidally induced
radial-velocity variations are almost exclusively caused by the tide
involved in the resonance and therefore exhibit a very sinusoidal-like
behaviour. Outside resonance, the radial-velocity variations are less
regular functions of the orbital phase. 

For conclusion, we applied our results to the slowly pulsating B star
HD\,177863 and showed the possible resonant excitation of a high-order
second-degree sectoral $g^+$-mode in this star.

\begin{acknowledgements}
The authors express their sincere thanks to Dr.\ A. Claret for
providing them with theoretical stellar models and to an anonymous
referee whose valuable comments led to an improvement of the
paper. Bart Willems acknowledges the financial support of PPARC grant
PPA/G/S/1999/00127. 
\end{acknowledgements}

\aareferences

\appendix

\section{The functions $f_{\xi,\ell}(u)$ and $f_{\eta,\ell}(u)$}

\label{fxe}

The functions $f_{\xi,\ell}(u)$ and $f_{\eta,\ell}(u)$ defined by
Eqs.\ (\ref{v:10a}) and (\ref{v:10b}) render the influence of the
adopted limb-darkening law on the contributions of the radial and the
transverse component of the tidal displacement to the tidally induced
radial-velocity variations. In the case of the limb-darkening law
given by Eq.\ (\ref{v:7}), the functions $f_{\xi,\ell}(u)$ and
$f_{\eta,\ell}(u)$ take the form  
\begin{equation}
f_{\xi,\ell}(u) = {6 \over {3-u}}\, \left[ 
  (1-u)\, \gamma_{\ell,2} + u\, \gamma_{\ell,3} \right], \label{f:1}
\end{equation}
\begin{eqnarray}
\lefteqn{f_{\eta,\ell}(u) = {6 \over {3-u}}\, \ell\, 
  \left[ (1-u) \left( \gamma_{\ell,2} - \gamma_{\ell-1,1} 
  \right) \right. }  \nonumber \\
 & & \left. + u \left( \gamma_{\ell,3} - 
  \gamma_{\ell-1,2} \right) \right], \label{f:2}
\end{eqnarray}
where we have introduced the abbreviation
\begin{equation}
\gamma_{\ell,n} \equiv \int_0^1 P_\ell(x)\, x^n\, dx.
  \label{f:3}
\end{equation}
For non-negative values of $n$, the integral in the right-hand member
of Eq.\ (\ref{f:3}) yields
\begin{equation}
\gamma_{\ell,n}= {{\pi^{1/2}\, \Gamma(n+1)} \over {2^{n+1}\, 
  \Gamma[(n-\ell+2)/2] \Gamma[(n+\ell+3)/2]}}, 
  \label{f:3b}
\end{equation}
[\citeauthor{Abr1965} \citeyear{Abr1965}, Eq. (8.14.15)].

In the particular case where $\ell=2$, the functions $f_{\xi,2}(u)$ and
$f_{\eta,2}(u)$ take the form   
\begin{equation}
f_{\xi,2}(u) = {1 \over {20}}\, {{16-u} \over {3-u}},  \label{f:4}
\end{equation}
\begin{equation}
f_{\eta,2}(u) = - {3 \over {10}}\, {{8-3\,u} \over {3-u}}.  \label{f:5}
\end{equation}

\end{document}